\documentclass[preprint,superscriptaddress,amsmath,amssymb,aps,prl]{revtex4-1}
\usepackage{color, amssymb, graphicx, amsfonts,epstopdf }
\usepackage{multirow,amssymb,amsbsy,amsmath,epstopdf}
\usepackage{hyperref}% add hypertext capabilities
\usepackage{graphicx}
\usepackage{verbatim}
\newcommand{\bra}[1]{\ensuremath{\left\langle #1\right|}}
\newcommand{\ket}[1]{\ensuremath{\left|#1\right\rangle}}
\newcommand{\proj}[1]{\ensuremath{\left|#1\right\rangle\!\!\left\langle #1\right|}}

\newcommand{\braket}[2]{\ensuremath{\left\langle #1|#2\right\rangle}}
\newcommand{\expval}[1]{\ensuremath{\langle #1 \rangle}}

\newcommand{\abs}[1]{\ensuremath{\left| #1 \right|}}

\begin{document}

\title{Direct measurement of a nonlocal entangled quantum state}
\author{Wei-Wei Pan}
%\email{xuxiaoye@ustc.edu.cn}
\author{Xiao-Ye Xu}
\affiliation{CAS Key Laboratory of Quantum Information, University of Science and Technology of China, Hefei 230026, People's Republic of China}
\affiliation{Synergetic Innovation Center of Quantum Information and Quantum Physics, University of Science and Technology of China, Hefei 230026, People's Republic of China}
%%%%%%%%%%%%%%%%%%%%%%%%%%%%%%%%%%%%%%%%%%%%%%%%%%%%%%%%%%%%%%%%%%%%%%%%%%%%%%
\author{Yaron Kedem}
\email{yaron.kedem@fysik.su.se}
\affiliation{Department of Physics, Stockholm University, AlbaNova University Center, 106 91 Stockholm, Sweden}
%%%%%%%%%%%%%%%%%%%%%%%%%%%%%%%%%%%%%%%%%%%%%%%%%%%%%%%%%%%%%%%%%%%%%%%%%%%%%%
\author{Qin-Qin Wang}
\author{Zhe Chen}
\author{Munsif Jan}
\author{Kai Sun}
\author{Jin-Shi Xu}
\author{Yong-Jian Han}
%\email{smhan@ustc.edu.cn}
\author{Chuan-Feng Li}
\email{cfli@ustc.edu.cn}
\author{Guang-Can Guo}
\affiliation{CAS Key Laboratory of Quantum Information, University of Science and Technology of China, Hefei 230026, People's Republic of China}
\affiliation{Synergetic Innovation Center of Quantum Information and Quantum Physics, University of Science and Technology of China, Hefei 230026, People's Republic of China}

\date{\today}
\begin{abstract}
Entanglement and wave function description are two of the core concepts that make quantum mechanics such a unique theory. A method to directly measure the wave function, using Weak Values, was demonstrated by Lundeen et al., Nature \textbf{474}, 188(2011). However it is not applicable to a scenario of two disjoint systems, where nonlocal entanglement can be a crucial element, since that requires obtaining Weak Values of nonlocal observables. Here, for the first time, we propose a method to directly measure a nonlocal wave function of a bipartite system, using Modular Values. The method is experimentally implemented for a photon pair in a hyper-entangled state, i.e. entangled both in polarization and momentum degrees of freedom.
\end{abstract}

\maketitle
A wavefunction description plays an important role in %the Copenhagen interpretation of quantum mechanics
quantum theory, while its objective reality gives rise to a century-long debate\,\cite{Bohr1935,Born1955,Born1927,Colbeck2012,Pusey2012,Ringbauer2015}. Using the technique of Weak Measurement that enables one to obtain the Weak Value of a pre- and post-selected quantum system\,\cite{AAV,AV90}, a method of directly measuring the complex wavefunction of single photons was experimentally demonstrated recently\,\cite{Lundeen2011}. %Recently, a complex wavefunction was measured directly\,\cite{Lundeen2011} using the technique of Weak Measurements\,\cite{AAV} which enables one to obtain the Weak Value of a pre- and post-selected system\,\cite{AV90}.
%It was generalized to mixed states \cite{general,density}, it was used to characterize polarization \cite{polar}, and orbital angular momentum \cite{27dim} states, to determine states of immense dimension \cite{compres} and to measure incompatible observables \cite{incomp}. 
The technique was subsequently extended to discrete two\,\cite{polar} or high dimension\,\cite{27dim,compres,Shi2015} quantum systems and even mixed states\,\cite{general,density}. The method was found to have deep connections to the phase space distributions\,\cite{Fischbach2012,Maurice2012} and sequential measurements\,\cite{Antonio2013a,Antonio2013b,incomp}. It was developed to with strong measurements\,\cite{Zou2015,strong,strong2} and further applied to measure matter waves\,\cite{Denkmayr2017,Denkmayr2018}.
%The technique was subsequently extended to discrete systems\,\cite{polar,27dim,compres,Shi2015} and mixed states\,\cite{general,density}. The method was found to have deep connections to phase space distributions\,\cite{Fischbach2012,Maurice2012} and sequential measurements\,\cite{Antonio2013a,Antonio2013b,incomp}. It was also applied to measure matter waves\,\cite{Denkmayr2017,Denkmayr2018}.
%
%The technique was subsequently extended to characterize discrete two\,\cite{polar} or high dimension\,\cite{27dim,compres} quantum state, further generalized to mixed states\,\cite{general,density}, and even developed to measure matter waves\,\cite{}. The method was found to have deep connections to the phase space distributions\,\cite{} and sequential measurements\,\cite{incomp}. 
%
In none of these tasks\,\cite{Dressel2014,Magana2016}, the measured wavefunction could be related to two disjoint systems and thus could not represent nonlocal entanglement. Here, for the first time, we show a direct measurement of a wavefunction with nonlocal entanglement. We achieve this by using Modular Values\,\cite{modular} which enable one to obtain the Weak Value of a (nonlocal) product of observables.

A general wavefunction $\ket{\Psi}$ can be written using a basis $\ket{n}$ as $\ket{\Psi} = \sum_n \Psi_n \ket{n}$, where $\Psi_n$ are complex amplitudes. A projective measurement of $\ket{n}$ would yield only $\abs{\Psi_n}^2$, and not any phase information, so it was a surprise when Lundeen {\it et. al.}\cite{Lundeen2011} showed that using Weak Values one can directly measure both the real and imaginary parts of $\Psi_n$. A Weak Value of an observable $O$, on a system that is prepared in a state $\ket{\psi}$  and postselected to a state $\ket{\phi}$, is given by $O_w = \bra{\phi} O \ket{\psi} / \braket{\phi}{\psi}$. It is a complex quantity, in contrast to the expectation value or any of the eigenvalues, which are always real. The Weak Value of a projection operator $P_n = \proj{n}$ with a postselection on uniform superposition $\ket{\phi} \propto \sum_n \ket{n}$ yields the complex amplitudes $\psi_n \propto \left(P_n\right)_w$.
The standard technique to obtain a Weak Value, known as a Weak Measurement, is via an interaction described by the evolution operator $U_I = e^{- i g O p}$, where $g \ll 1$ is a dimensionless coupling constant and $p$ is an operator on a meter. After the interaction and postselection on the system the expectation value $\expval{p}$ and $\expval{x}$, with $x$ being an operator conjugate to $p$, will change according to $\delta p = 2 g  \left( \expval{p^2} -\expval{p}^2 \right) \Im \left[ O_w \right]$ and $\delta x = g  \Re \left[ O_w \right]$, respectively \cite{snr}.

\begin{figure}
    \centering
    \includegraphics[width=6.263cm]{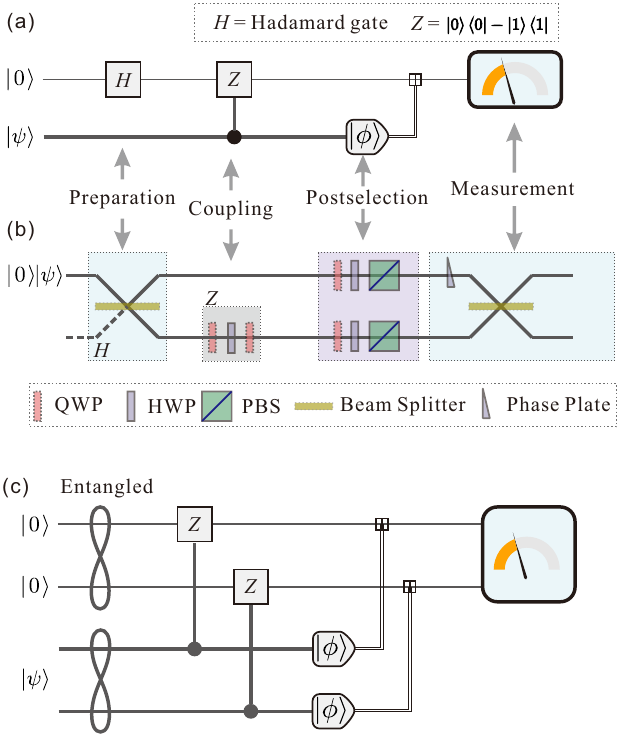}
 \caption{Measurement of a pre- and postselected system, using qubit meters (a) A logic circuit diagram is showing a single qubit meter, prepared in a state $\ket{\uparrow_x} = (\ket{0} + \ket{1})/\sqrt{2}$, and a single system prepared in a state $\ket{\psi}$. Then, a controlled gate modify the state of the meter, depending on the state of the system and finally, if the system is found in a state $\ket{\phi}$, the meter is read out. (b) Experimental scheme to realize the circuit in (a) using the path of a photon as a meter and its polarization as the system. (c) The logic circuit diagram for the case of a bipartite system (two spatiality separated qubits) is composed from similar components to the single qubit case shown in (a) and (b). The main difference is that the system can be initially entangled and thus the meter might have to be entangled as well.}    \label{fig:LogicDiagram}
\end{figure}

Consider the case that our system is composed of two subsystems, $A$ and $B$, placed at different locations. The complete Hilbert space is a tensor product of the two local Hilbert spaces $\mathcal{H} = \mathcal{H}_A \otimes \mathcal{H}_B $. One can define local bases for each subsystem $\ket{j}_A \in  \mathcal{H}_A, \ket{l}_B  \in  \mathcal{H}_B$ such that the basis for the complete system is a tensor product of local states $\ket{n}=\ket{j,l} = \ket{j}_A\ket{l}_B$, with amplitudes $\Psi_{j,l}$ given by the Weak Value of projection operators $P_{j,l} = P^A_jP^B_l$. However, the interaction $U_I = e^{- i g \left(P_{j,l} \right) p}$, which is needed in the standard scheme to obtain $\left(P_{j,l}\right)_w$, is not physical, regardless of what $p$ might be, since it requires a nonlocal Hamiltonian $H \propto P^A_j P^B_l$. Such a Hamiltonian implies an instantaneous interaction between distant locations. Thus, the method described above cannot be applied to this case and it seems that one cannot directly measure a nonlocal wavefunction. This implies a major flaw in the implication of the direct measurement technique. Apart from the ongoing discussing regarding the efficiency of the method, this flaw is related to the fundamental aspects of the idea and can render it useless for the most interesting cases. Without the locality restriction one can adopt a realistic description and thus the wavefunction is redundant. Any potential application of the method would also be highly limited due to the pivotal role of entanglement in many quantum protocols. Here we show that this is not the case by introducing a new method, and experimentally demonstrating it, where the Weak Values are replaced, or rather augmented, by Modular Values\cite{modular}.

Modular Values were introduced as an explanation of an experiment demonstrating the Hardy paradox\cite{hardy,hardyExp}, which involved the Weak Value of a product, and as a method to obtain Weak Values using strong measurement. Later on it extended theoretically in several ways \cite{geoTheo} %\cite{char,geoTheo,enlarged,genMod} 
and also implemented experimentally \cite{geoExp}.

Since the (nonlocal) observable we are interested in $P_{l,j}$ is a product of two (local) observables, the problem boils down to obtaining the Weak Value of a product of observables. This task cannot be done using the standard Weak Measurement technique, where the meter evolves according to a Hamiltonian in which the observable on the system is replaced by its Weak Value $O \rightarrow O_w$. A way to achieve this task was initially suggested in\,\cite{Resch} and later realized \,\cite{lundeen2009}. Their method relies on a second order term, while still requiring the interaction to be weak. A product of $N$ observables would be obtained from the $N$th order, so the scalability of this method poses significant challenges. The method we use, based on Modular Values, has the additional benefit of allowing one to obtain Weak Values using strong measurement. In \cite{modular}, it was shown that a qubit meter, interacting via an observable $O$ on a pre- and postselected system, evolves according to the Modular Value, given by
$O_m = \bra{\phi} e^{-i g O }\ket{\psi} / \braket{\phi}{\psi}$, where $g$ is coupling constant of arbitrary size.
When the relevant observable is a projection, the Modular Value has a close connection to the Weak Value $\left(P_n\right)_m = 1 + s \left(P_n\right)_w$, with $s = e^{-i g  }-1$. We set $s = -2$ which corresponds to a standard experimental setting. In the case of two commuting projectors we have
\begin{align} \label{psi}
\Psi_{j,l}&\propto \left(P^A_jP^B_l\right)_w \nonumber \\&= s^{-2} \left[ \left(P^A_j + P^B_l\right)_m - \left(P^A_j\right)_m - \left(P^B_l\right)_m +1\right],
 \end{align}
 where for any single projector on a subsystem there is an implicit tensor product with an identity operator on the other subsystem. The first expression in Eq.\,(\ref{psi}) is implied by the original method\,\cite{Lundeen2011} while the second expression comes directly from the definition of the Modular Value. While qubit meters are typically used to obtain Modular Values, the projection observable $P_j$ can pertain to a continuous variable such as the position or velocity of a particle \cite{steer,Xiao2019}, with the indices $j,l$ in Eq.\,(\ref{psi}) denoting the continuous property. Thus, the problem of measuring directly a nonlocal wavefunction is mapped to directly measuring the Modular Values of observables such as $P^A_j + P^B_l$, which we now show how to accomplish, using an entangled meter.

Since the meter should interact with both subsystems, it should also consists of two parts, even though in principle one can also have a single meter and move it to each location of the subsystems. After each part interacts with one subsystem, and the system is postselected, the Modular Value can be extracted from the final state of the meter by tomography \cite{modular}. The tomography in the last step can be replaced by a more direct method by setting the meter in an initial state $\ket{\Psi^m_I} =  \left( \ket{\uparrow \downarrow} + \epsilon \ket{\downarrow \uparrow} \right)/ \sqrt{1+\epsilon^2}$, where the first (second) arrow refer to the part of the meter interacting with subsystem $A(B)$ and $\epsilon \ll 1$. Then, after an interaction $U_I = e^{- i g \left(P^A_j P^A_\downarrow + P^B_l P^B_\uparrow\right)}$ with $P^A_\downarrow$ ($P^B_\uparrow$) a projection on the part $A(B)$ of the meter to the state $\ket{\downarrow}$ ($\ket{\uparrow}$), the probabilities $\mathcal{P}_1$ and $\mathcal{P}_2$ of finding the meter in the states $\ket{1} =  \left( \ket{\uparrow \downarrow} + \ket{\downarrow \uparrow} \right)/\sqrt{2}$ and $\ket{2} =  \left( \ket{\uparrow \downarrow} + i \ket{\downarrow \uparrow} \right)/\sqrt{2}$, respectively, are given by
 \begin{align}\label{finalStateR}
\mathcal{P}_1 &=  {\frac{1}{2}} + \epsilon \Re   \left(P^A_j + P^B_l\right)_m+  O(\epsilon^2) ,\\
\mathcal{P}_2 &=  {\frac{1}{2}} + \epsilon \Im   \left(P^A_j + P^B_l\right)_m +  O(\epsilon^2). \label{finalStateI}
  \end{align}
Thus, to the first order in $\epsilon$, the readout in certain detectors will be given by the real and imaginary part of the relevant Modular Value.   
To obtain $\left(P^A_j\right)_m$ and $\left(P^B_l\right)_m$ one can set the meter initially in a product state and look at probability of finding states, similar to $\ket{1}$ and  $\ket{2}$, for each part of the meter separately. Alternatively, one can replace in the interaction $U_I$, a projector on one part with an identity operator (for more details see the supplementary information).

\begin{figure}
    \centering
    \includegraphics[width=0.49\textwidth]{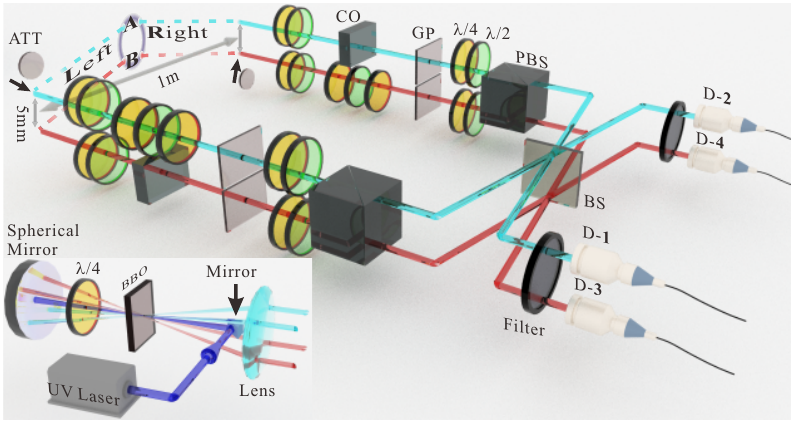}
    \caption{Experimental setup. The inset is showing a $\beta$-BaB$_2$O$_4$ (BBO) crystal pumped by a vertically polarized ultraviolet laser. A spherical mirror reflects the photon pairs, which are emitted due to a degenerate spontaneous parametric down conversion, and the pump beam to a second pass through the crystal, after passing a quarter-wave plate (QWP or $\lambda$/4) with its optical axis orienting at 45$^\circ$. Choosing four points in the entanglement ring yields a photon pair entangled both in path and polarization, which enters the main setup shown in the main panel. The up path (green) stands for the first photon (labeled $A$) while the bottom path (red) for the second one (labeled $B$), they are vertically separated around 5\,mm. For each photon, they have two spatially separated path modes (left and right with around one meter from each other) and two polarization modes (horizontal and vertical), the former is taken as the meter and the latter as the system. For preparing the initial meter state, we use two neutral optical attenuators (ATTs) to precisely reducing the relative intensity of relevant modes. Half- and quarter- wave plates are used to manipulate the polarization of each path independently and glass compensators (COs) are used to offset the phase difference. The two paths of each photon are combined in one unpolarized beam splitters (BS). Glass plates (GPs) are used to introduce a controllable phase so different path states can be measured. Polarized beam splitter (PBS) postselect the final polarization state. After spectrum filters, the photons are collected by four single mode fibers (SMFs) and guided to four avalanche single-photon detectors (D-1,2,3,4).}
    \label{fig:Setup}
\end{figure}

Note that $\epsilon \ll 1$ does not imply the procedure is a Weak Measurement in the sense that the interaction parameter $g$ does not have to be small. One can set $\epsilon$ to be large as well and still reconstruct Weak Values from Eq.\,(\ref{psi}), which means using strong measurements. Indeed an alternative to the direct measurement technique, based on strong measurement, was theoretically proposed\,\cite{strong,Zou2015} and experimentally demonstrated\,\cite{strong2}. That technique could also be interpreted using Modular Values. In our scheme, we choose $\epsilon \ll 1$, so using Eq.\,(\ref{finalStateR}) and (\ref{finalStateI}), the complex amplitudes of a wavefunction with nonlocal entanglement, appear naturally in the measurement results.

The general scheme of our method is shown in Fig.\,\ref{fig:LogicDiagram} and the experimental setup is shown in Fig.\,\ref{fig:Setup}. More details are given in the supplementary. In the experiment we have measured the wavefunction of the polarization of two photons using their paths degree of freedom as a meter, such that the state $\ket{\uparrow}$ ($\ket{\downarrow}$) for a part of the meter implies that photon went through the right (left) arm. We used hyperentangled photon pairs %\,\cite{Kwiat1997} 
such that both the polarization and path states are entangled between the photons while there is no entanglement, initially, between the polarization and path\,\cite{Barbieri2005}. The probabilities in Eq.\,(\ref{finalStateR}) and (\ref{finalStateI}) are obtained by Franson interference. Since the initial state of the meter is entangled, we detect product states $\ket{\tilde{1}} =  \left( \ket{\uparrow } + \ket{ \downarrow} \right) \left( \ket{\uparrow } + \ket{ \downarrow} \right)/2$ and $\ket{\tilde{2}} =  \left( \ket{\uparrow } + i \ket{ \downarrow} \right) \left( \ket{\uparrow } + \ket{ \downarrow} \right)/2$ to obtain $\mathcal{P}_1$ and $\mathcal{P}_2$, respectively. 

\begin{figure}
    \centering
    \includegraphics[width=0.375\textwidth]{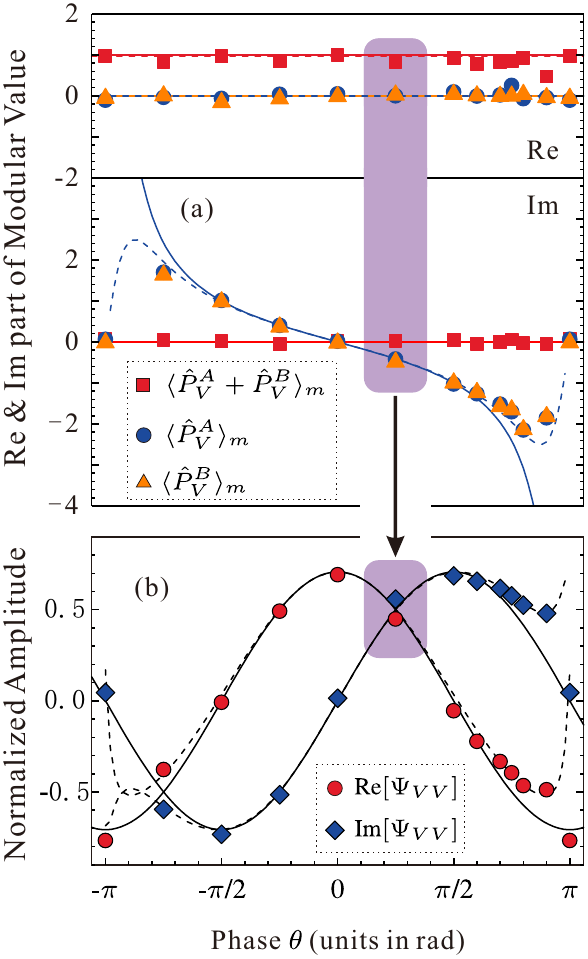}
    \caption{Demonstration of the method for the state $\frac{1}{\sqrt{2}}(\ket{HH} + e^{i\theta}\ket{VV})$. (a) Modular values (upper panel for real part and lower panel for imaginary part): To avoid clutter, we only show the corresponding values for the projector $P_V^A$, $P_V^B$ and $P_V^A + P_V^B$. (b) The normalized probability amplitude for the corresponding component $\Psi_{VV}$ of the state. While all the real part of modular values in (a) exhibit constant value, the oscillations in both the real and imaginary part of $\Psi_{VV}$ emerge when performing the normalization to the whole state. In both panels, solid lines are values derived from a perfectly prepared state, dashed lines are predicted results taking into account higher order of $\epsilon$ ($=0.2$ in our case) and markers are experimental results. The error bars (calculated through Monte Carlo simulations considering the counting noise) are smaller than the point size.}
    \label{fig:WeakValues}
\end{figure}

\begin{figure}
    \centering
    \includegraphics[width=8cm]{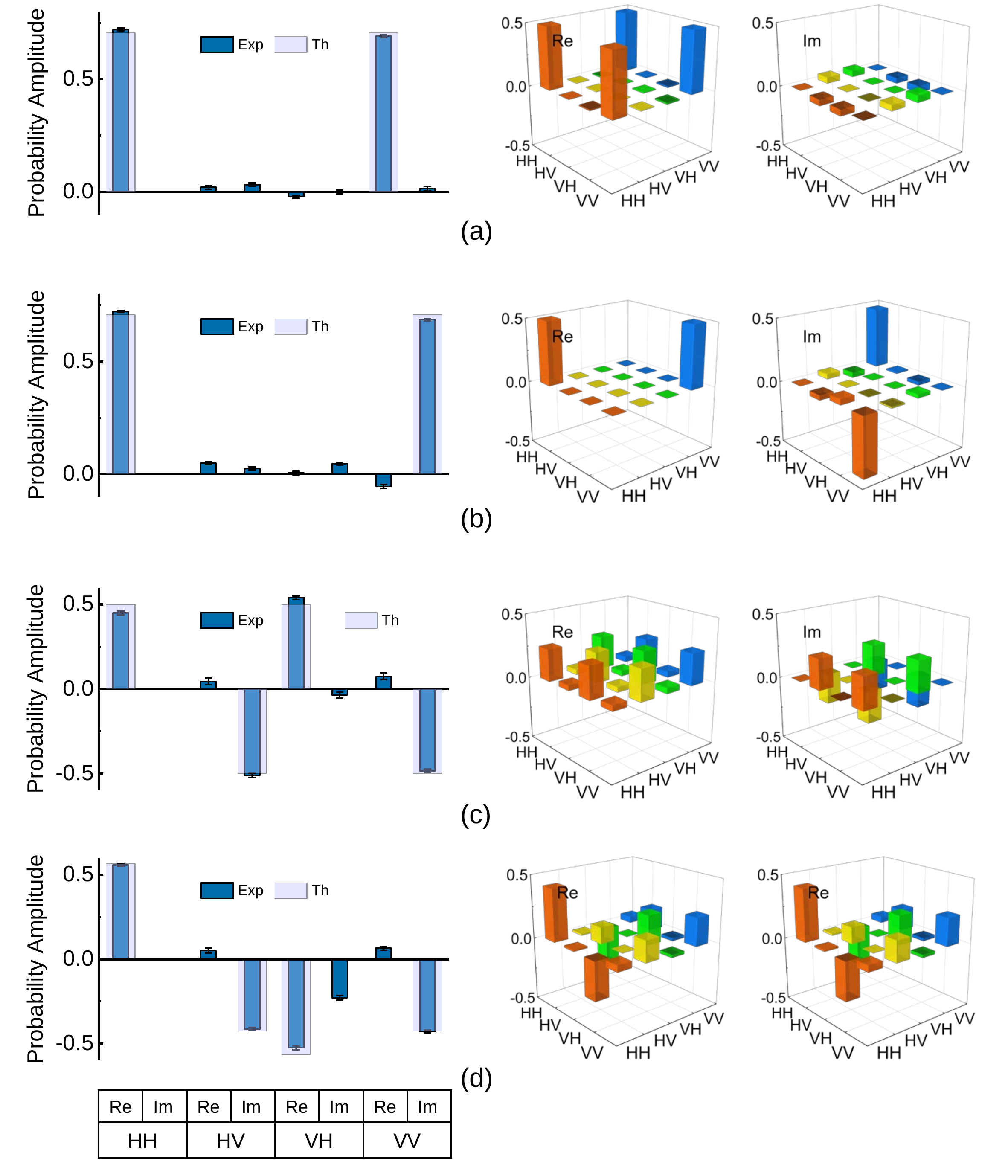}
   \caption{Experimentally measured probability amplitudes. The upper part of each panel shows the directly measured probability amplitudes (Exp). The theoretical values (Th) refer to the state in case of an ideal preparation, which in each panel is given by (a) $\ket{\Psi} =(\ket{HH} + \ket{VV})/\sqrt{2}$ (b) $\ket{\Psi} =(\ket{HH} + i \ket{VV})/\sqrt{2}$ (c) $\ket{\Psi} =(\ket{H} + \ket{V}) (\ket{H} - i \ket{V})/2$ and (d) $[0.8\ket{HH}-0.6i\ket{HV}-0.8\ket{VH}-0.6i\ket{VV}]/\sqrt{2}$. The lower part of each panel shows the reconstructed density matrix (colored bars) for each state, obtained by standard tomography. Errors are estimated through Monte Carlo simulations considering the counting noise. }
   \label{fig:Coefficient}
\end{figure}

We start by demonstrating our method for a single component of the state. We prepare the system in a maximally entangled state $\frac{1}{\sqrt{2}}(\ket{HH} + e^{i\theta}\ket{VV})$ with an adjustable phase $\theta\in\{-\pi,\pi\}$ and postselect the system to $(\ket{H}+\ket{V})(\ket{H}+\ket{V})/2$, where $\ket{H}$ and $\ket{V}$ denotes the horizontal and vertical polarization, respectively. The relevant Modular values and the probability amplitude are shown in Fig.\,\ref{fig:WeakValues}. The method %  using Eq.\,(\ref{psi}),  (\ref{finalStateR}) and (\ref{finalStateI}) 
yields the expected values for $\abs{\theta} \lesssim \pi/2$. For $\abs{\theta} \simeq \pi$, where the initial state is nearly orthogonal to the postselection, the Modular Value diverges, as does the Weak Value, while the probability amplitude does not. The relation in Eq.\,(\ref{psi}) still holds since the proportionality constant in the first expression vanishes accordingly. However, higher orders in Eq.\,(\ref{finalStateR}) and (\ref{finalStateI}) cannot be neglected anymore. This problem can be solved by choosing a different postselection, for example $(\ket{H}+\ket{V})(\ket{H}-\ket{V})/2$. Another solution would be to make $\epsilon$ smaller if the interferometer is sufficiently ideal with high interference visibility.      

In Fig.\ref{fig:Coefficient} we present the complete probability amplitudes which are measured using our method, for a number of states. In our case, the three modular values used to measure the single component, as shown in Fig.\,\ref{fig:WeakValues} are enough to obtain all the components, since $P_H = I-P_V$, with $I$ the identity matrix, and $ \left(c I +O\right)_m = s^c \left(O\right)_m$, for any observable $O$ and number $c$. A system of dimension $n$ would have $n-1$ independent projectors, so for a general bipartite system, composed of subsystems of dimensions $m$ and $n$, one needs $(m-1)+(n-1)$ single-system Modular Values, and $(m-1)\times (n-1)$ two-system combinations, i.e. products of projectors on the separate systems. Adding these numbers and multiplying by 2, for real and imaginary parts, yield exactly the number of independent parameters $2 m \times n -2$ in a state of dimension $n \times m$. For comparison we show in Fig.\ref{fig:Coefficient} the reconstructed density matrices, obtained by standard tomography, requiring 16 different measurements for each state. Our current work is limited to pure states and reconstructing mixed states will necessarily be more resource intensive.

In addition to the main result, our method illuminate an aspect of the direct measurement technique, which is sometimes overlooked or misunderstood. It demonstrate that essentially the ability to obtain the probability amplitude is due to Weak Values rather than Weak Measurements. It is the postselection that enables us to achieve the task and not necessarily a small interaction strength. While this distinction could be infer directly from the theoretical derivation, supporting it by experimental results, can help clarify this issue, as well as the discussion regarding the efficiency of various techniques, especially with regards to the supposed benefits of small interactions.

More importantly, extending the method of direct measurement to scenarios having nonlocal entanglement, will allow using it to study, theoretically and experimentally, many ideas, such as steering, quantum discord, entanglement entropy, etc. The extended method could be incorporated into quantum protocols for which the original method was not applicable due to the locality constraint on the measured wavefunction. Since it is not yet clear where one would use the original method, and how, we can only speculate regarding concrete applications of our method. It is possible that some future technology would require obtaining the wave function of a pure state, which is non locally entangled. The context can be information transfer, executing a distributed computational task, cryptography protocols, etc. Then, one might find that our method is required for an efficient implementation of such technology.

In conclusion, we have experimentally demonstrated a direct measurement of nonlocal wavefunctions for the first time. The task is achieved by using Modular Values of a sum of observables which yield the Weak Values of nonlocal observables. The method sheds new light on the previous technique and extends it to be applicable for an important scenario: the existence of nonlocal entanglement. 
%Further works,  matter system, solid state system, superconductor, mixed state (sequential weak values) 
We anticipate that our results can inspire the direct measurement of multipartite states in some other quantum systems where weak values are accessible at present, such as atoms\,\cite{shomroni2013}, neutrons\,\cite{sponar2015}, superconductor\,\cite{marian2016}, etc. Introducing sequential measurements\,\cite{Antonio2013a,Antonio2013b,incomp} into our method and developing it for directly measuring mixed states should be worthy of further studies. The simplicity of the theoretical derivation, and the demonstrated feasibility of the experimental technique, can make the new method a powerful tool for studying the nature of quantum mechanics and harnessing it. 

\begin{acknowledgments}
%We acknowledge the insightful comments of the anonymous referees.
Wei-Wei Pan and Xiao-Ye Xu contributed equally to this work. 
This work was supported by 
National Key Research and Development Program of China (Nos.\,2017YFA0304100, 2016YFA0302700), 
the National Natural Science Foundation of China (Nos.
%Yong-Jian Han
\,11874343, 
%Xiao-Ye Xu
61805228, 
%Chuan-Feng Li
11774335, 11821404, 
%Jin-Shi Xu
61725504), 
Key Research Program of Frontier Sciences, CAS (No.\,QYZDY-SSW-SLH003), 
Science Foundation of the CAS (No. ZDRW-XH-2019-1),
the Fundamental Research Funds for the Central Universities (No.\,WK2470000026),  %
the National Postdoctoral Program for Innovative Talents (No.\,BX201600146), 
China Postdoctoral Science Foundation (No.\,2017M612073) and 
Anhui Initiative in Quantum Information Technologies (No.\,AHY020100, AHY060300).
\end{acknowledgments}

\bibliographystyle{apsrev4-1PRX}
\bibliography{DirectEntag}

\clearpage
\newpage
\setcounter{page}{1}
\appendix 
\setcounter{equation}{0}
\renewcommand{\theequation}{S\arabic{equation}}
\section*{Supplementary Information}

\vspace{1em}
\textbf{Hyper-entangled photon source.} Hyper-entangled photon pairs are useful in quantum information, for its convenience in implementing the inner-dimensions coupling, avoiding the challenge in realizing photon-photon couplings in multiphoton schemes. Here we follow the confocal structure using type-\uppercase\expandafter{\romannumeral1} SPDC for preparing the polarization-path hyper-entangled state. The wavelength of pumper laser is 406.7\,nm and vertically polarized. The thickness of BBO crystal is 0.5 mm. It is cut at 29.11$^\circ$ for degenerate type-\uppercase\expandafter{\romannumeral1} SPDC. We use four spectrum filters centered at 813.4\,nm with bandwidth 3\,nm to extract both the signal and idle photons. The confocal structure makes the photons outputting in a ring shape. As shown in Fig\,\ref{fig:Setup} of the main text, four points in the ring are preselected for our experiment, two on the top colored green are adopted as photon-$A$ and two on the bottom colored red as photon-$B$. Each photon has two degrees of freedom, one is the spatial (left $\ket{\downarrow}$ and right $\ket{\uparrow}$ with around one meter from each other) taken as the meter and the other is the polarization (horizontal $\ket{H}$ and vertical $\ket{V}$) taken as the system. We use four single mode fibres to collect the photons, two (D-1, D-2) for photon-$A$ and two (D-3, D-4) for photon-$B$. After precise compensation of time delay, the photons are prepared in a hyper-entangled state, which takes the form
\begin{equation}\label{Eq.S1}
    \ket{\Psi^{ms}} = \tfrac{1}{\sqrt{2}}(\ket{\uparrow\downarrow}+\ket{\downarrow\uparrow})^m\otimes\tfrac{1}{\sqrt{2}}(\ket{HH}+\ket{VV})^s.
\end{equation}

\vspace{1em}
\textbf{Initialization of the system and meter.} Initially, the photon pairs are prepared in a hyper-entangled state such that both the polarization and path states are maximally entangled between the photons while there is no entanglement between the two degrees of freedom. Starting from the maximal entangled state given in Eq.\,\ref{Eq.S1}, the system can then be prepared to any required form (maximally entangled) by using the wave plate set of HWP-QWP in the beginning of each arm. At this stage, the optical axis of the QWP inside the photon source (see the inset in Fig.\,\ref{fig:Setup}) is oriented at 45$^\circ$ (by 0$^\circ$ we means the axis is parallel to the table). For preparing the system in a product state, in addition to wave plate sets used before we also need to set the QWP inside the photon source to 0$^\circ$. For the general state demonstrated in our experiment, a careful configuration of the optical axes of the QWP and the wave plate sets is needed. In our scheme, a small value of $\epsilon$ is needed. Since the meter is not initially entangled with the system, we can implement it by inserting neutral optical attenuators, which can reduce the optical intensity independent of its polarization, into the relevant path modes, i.e., left for photon-$A$ and right for photon-$B$. By precisely reducing the intensity of the path mode \ket{\downarrow^A\uparrow^B}, we can set the meter to $\Psi^m_I = (\ket{\uparrow\downarrow}+\epsilon\ket{\downarrow\uparrow})/\sqrt{1+\epsilon^2}$ with $\epsilon$ equaling 0.2 in our work.  

\vspace{1em}
\textbf{Measuring the Modular Values.} Using the setup shown in the main text, and by inserting different configurations of wave plates, we implement the following time-evolution operators:
\begin{align}
&U_1=e^{-i\pi(P^A_\downarrow P^A_j+P^B_\uparrow P^B_l)},\\
&U_2=e^{-i\pi P^A_\downarrow P^A_j},\\
&U_3= e^{-i \pi P^B_\uparrow P^B_l},
\end{align}
with $j,l\in\{H,V\}$ and $P$ as a projection operator. An implicit tensor product with the identity operator on any other part of the system is assumed. The initial state of the meter is $\ket{\Psi^m_I}$, as defined in the main text  with $\epsilon=0.2$. After the evolution, under $U_1$, $U_2$ or $U_3$, and postselection, the state of the meter is, respectively:
\begin{align}
\ket{\Psi^m_F}_1&=\mathcal{N}\left[ \epsilon\left( P^A_j+P^B_l\right)_m\ket{\downarrow\uparrow}+\ket{\uparrow\downarrow}\right],\\
\ket{\Psi^m_F}_2&=\mathcal{N}\left[ \epsilon\left( P^A_j\right)_m\ket{\downarrow\uparrow}+\ket{\uparrow\downarrow}\right],\\
\ket{\Psi^m_F}_3&=\mathcal{N}\left[ \epsilon\left( P^B_l\right)_m\ket{\downarrow\uparrow}+\ket{\uparrow\downarrow}\right],
\end{align}
where $\mathcal{N} = 1- O(\epsilon^2)$ is a normalization factor. Projecting these states $\ket{\Psi^m_F}_{(1-3)}$ to $\ket{1}$ and $\ket{2}$ defined in the main text will yield probabilities $\mathcal{P}_1$ and $\mathcal{P}_2$ in terms of the corresponding Modular Values, as given in Eq.\,(2) and (3). In more detail, considering the Modular Value of the sum of two projectors $\left( P^A_j+P^B_l\right)_m$, we have
\begin{align}
    \mathcal{P}_1 &= |\langle 1 |\Psi_F^m\rangle_1|^2 \approx \epsilon\Re[(P_j^A + P_l^B)_m]+\tfrac{1}{2},\\
    \mathcal{P}_2 &= |\langle 2 |\Psi_F^m\rangle_1|^2 \approx \epsilon\Im[(P_j^A + P_l^B)_m]+\tfrac{1}{2}.
\end{align}
Similarly, we can get the Modular Value of the two single projectors $\left( P^A_j\right)_m$ and $\left(P^B_l\right)_m$. As the final meter states are entangled, especially they do not contain the components of $\ket{\uparrow\uparrow}$ and $\ket{\downarrow\downarrow}$, the same projection probability can be obtained by using $\ket{\tilde{1}}$ and $\ket{\tilde{2}}$ instead of $\ket{1}$ and $\ket{2}$. We have adopted the entangled meter at the beginning, our projection measurement works for the Franson interference occurring between the two spatially entangled photons. At last, with the experimentally measured Modular Values in hand and the formula in Eq.\,\ref{psi}, we can easily calculate the weak values of joint projector $(P_j^AP_l^B)_w$.

\textbf{Relationship between the complex probability amplitude and the joint weak value.} Supposing we have a system composed of two qubits $A$ and $B$ in a general normalized state
\begin{equation}
    \ket{\Psi^s} = \sum_{j,l}\Psi_{j,l}\ket{j}_A\ket{l}_B,
\end{equation}
where $\sum_{j,l} |\Psi_{j,l}|^2 \equiv 1$. The task is to measure the complex amplitude directly from the projection probabilities. Additionally, supposing we can extract the weak values of joint projectors $(P_j^AP_l^B)_w$, along its definition
\begin{equation}
    (P_j^AP_l^B)_w = \frac{\bra{\Phi^s}P_j^AP_l^B\ket{\Psi^s}}{\bra{\Phi^s}\Psi^s\rangle}.
\end{equation}
When choosing $\ket{\Phi^s} = \ket{++}$ with $\ket{+} = \tfrac{1}{\sqrt{2}}(\ket{H} + \ket{V})$, we can directly get the relationship between the complex probability amplitude and the joint weak values
\begin{equation}
    \Psi_{j,l} \propto (P_j^AP_l^B)_w,
\end{equation}
with a common constant coefficient to all the four components. When we have the experimental results of joint weak values, i.e., $(P_j^AP_l^B)_w^\text{exp}$, the complex amplitudes measured in experiment then read 
\begin{equation}
    \Psi_{j,l}^\text{exp} = \frac{(P_j^AP_l^B)_w^\text{exp}}{(P_H^AP_H^B)_w^\text{exp}\tilde{\mathcal{N}}}
\end{equation}
with the normalization coefficient
\begin{equation}
    \tilde{\mathcal{N}} = \sqrt{\sum_{j,l}\left |\frac{(P_j^AP_l^B)_w^\text{exp}}{(P_H^AP_H^B)_w^\text{exp}}\right |^2}.
\end{equation} %Note that $\tilde{\mathcal{N}}$ is a complex number in general and is dependent on the system's initial state and the postseleted state. 
That is to say, with all the joint weak values in hand, the full state (wave function) can then be obtained directly by performing a simple normalization to all the components. For example, considering the case in Fig.\,3, we have $\Psi_{HH} = 1/\sqrt{2}$, $\Psi_{VV} = e^{i\theta}/\sqrt{2}$, and $\Psi_{HV} = \Psi_{VH} = 0$. Consequently, the expected joint weak values in experiment read $\overline{(P_H^AP_H^B)_w} = 1/(1+e^{i\theta})$, $\overline{(P_V^AP_V^B)_w} = e^{i\theta}/(1+e^{i\theta})$, and $\overline{(P_H^AP_V^B)_w} = \overline{(P_V^AP_H^B)_w} = 0$. Meanwhile, $\tilde{\mathcal{N}} = \sqrt{2}$. While the real part of $\overline{(P_V^AP_V^B)_w}$ takes a constant value independent of the parameter $\theta$, after performing the normalization procedure, the final normalized amplitude for the basis $\ket{VV}$ reads $e^{i\theta}/\sqrt{2} = (\cos{\theta} + i\sin{\theta})/\sqrt{2}$, whose real and imaginary parts both exhibit oscillations, as shown in Fig.\,3.

%$\tilde{\mathcal{N}} = \sqrt{2}/(1+e^{i\theta})$. For the amplitude $\Psi_{VV}$, while all the real parts of the modular values, which directly give the real parts of corresponding weak values, take constant values, the normalized amplitudes read $e^{i\theta} = \cos{\theta} + i\sin{\theta}$, whose real and imaginary parts still exhibit oscillations, as shown in Fig.\,3.

\end{document}